# Short Ballistic Josephson Coupling in Planar Graphene Junctions with Inhomogeneous Carrier Doping

Jinho Park[1], Jae Hyeong Lee[1], Gil-Ho Lee[1], Yositake Takane[2], Ken-Ichiro Imura[2], Takashi Taniguchi[3], Kenji Watanabe[3], Hu-Jong Lee[1]

[1]Department of Physics, Pohang University of Science and Technology, 37673, Korea
[2]Department of Quantum Matter, AdSM, Hiroshima University, 739-8530, Japan
[3]Advanced Materials Laboratory, National Institute for Materials Science, 305-0044, Japan

**Abstract**

We report on short ballistic (SB) Josephson coupling in junctions embedded in a planar heterostructure of graphene. Ballistic Josephson coupling is confirmed by the Fabry–Perot-type interference of the junction critical current $I_c$. An exceptionally large $I_c R_N$ product close to $2\Delta_0/e$ ( $R_N$ ; normal-state junction resistance, $\Delta_0$ ; zero-temperature superconducting energy gap) is an indication of SB strong Josephson coupling. $I_c$ shows a temperature dependence inconsistent with the conventional short-junction-like behavior; $I_c(T)$ curves deviate systematically from the standard Kulik–Omel'yanchuk prediction. We argue that this feature stems from the planar nature of graphene junction, which is susceptible to the effects of inhomogeneous carrier doping as well in graphene near the superconducting contacts.

Graphene shows a characteristic linear dispersion (a synthetic relativistic world) in momentum space, whereas its real-space nature is also very unique, realizing a hitherto unknown perfectly flat two-dimensional world [1-3]. The exotic proximity properties arising from this intrinsic linear dispersion in graphene have been studied with the graphene-superconductor heterostructures [4-8]. In particular, a planar graphene Josephson junction (pGJJ), formed by arranging two superconducting electrodes in close proximity on a graphene layer, have been widely adopted [7-12]. While these studies of pGJJs took good account of the linear dispersion, unusual characteristics of pGJJs reflecting the unique planar nature of graphene has not been fully paid attention to.

In a pGJJ the superconducting order of the two adjacent superconductors is coupled by forming the Andreev bound state (ABS) in the graphene insert. Short ($L < \xi$; $L =$ junction channel length, $\xi =$ superconducting coherence length) and ballistic ($L < l_{mfp}$; $l_{mfp} =$ mean free path) strong Josephson coupling, via a single ABS, is essential to realizing coherent quantum states. As a pGJJ allows gate tuning of the Josephson coupling strength, realizing short ballistic (SB) strong Josephson coupling in pGJJs would offer their versatile applications to quantum devices.

Recently, ballistic JJ characteristics have been reported in pGJJs prepared by hexagonal-boron-nitride (hBN) encapsulation of a graphene layer together with atomic edge contacts [13-17]. In addition, the SB Josephson characteristics have been reported in vertical [18,19] and planar [14] junctions. However, contrary to vertical graphene JJs, where the channel length is mainly governed by the atomic thickness of the graphene insert, the SB nature in pGJJs [14] has yet to be confirmed. The reported $I_c R_N$ product ($I_c$; junction critical current, $R_N$; normal-state junction resistance) is far smaller than the theoretical prediction and $I_c$ decays exponentially with increasing temperature near $T_c$ (see Fig. S5), a canonical long-junction ($L > \xi$) behavior. The $I_c R_N$ product represents the Josephson coupling strength, which constitutes the important figure of merit of a proximity JJ. Thus, realization of SB strong Josephson coupling still remains highly challenging in pGJJs.

Here, we report on the SB Josephson coupling in pGJJs with an Al superconducting contact. With a relatively small superconducting gap, Al provides a long superconducting coherence, which is advantageous for realizing a short junction ($L < \xi$). The value of $I_c R_N$ of pGJJs normalized by $\Delta_0/e$ in this study is much larger than the ones obtained previously in pGJJs [14] in which SB characteristics were claimed. The value of the $I_c R_N$ product in this study (~ $2\Delta_0/e$) almost reached the theoretically predicted value (~$2.4\Delta_0/e$) for an ideal SB proximity graphene JJ [20]. Especially, we focused on the $T$ dependence of $I_c$ to examine the SB nature of our pGJJs. For the $I_c(T)$ fit, we adopted the theoretical model [21,22] taking into account carrier doping near the graphene–superconductor interfaces. The two-dimensional nature of graphene significantly enhances the doping effect on $I_c(T)$ of a junction. $I_c(T)$ curves of our pGJJs show a near-perfect fit with this model, whereas they are in apparent discord with the conventional Kulik-Omel'yanchuk (KO) prediction [23,24], which is often used to examine the coupling characteristics of a proximity JJ.

We prepared a stacked hexagonal boron nitride (hBN)-graphene-hBN structure using a dry-transfer technique [25]. Superconducting contacts to the monolayer graphene were made at the junction edges, exposed by the $O_2/CF_4$ plasma etching of BN flakes and the graphene layer simultaneously. Edge contacts were made to the graphene layer [25] by in-situ electron-beam deposition of two Al/Ti bilayer (60/6 nm thick) superconducting electrodes. Two junctions were prepared simultaneously; SBJJ40-1 and SBJJ40-2 (right and left in Fig. 1(a)). The physical scale of the junctions was estimated by scanning electron microscopy; both had a common channel width of ~4.7 μm, and channel lengths of ~120 nm (SBJJ40-1) and ~220 nm (SBJJ40-2). The junction resistance was measured by a quasi-four-probe scheme, in which the contact resistance ($R_c$) of two graphene–Al interfaces was included (see Fig. 1(a)).

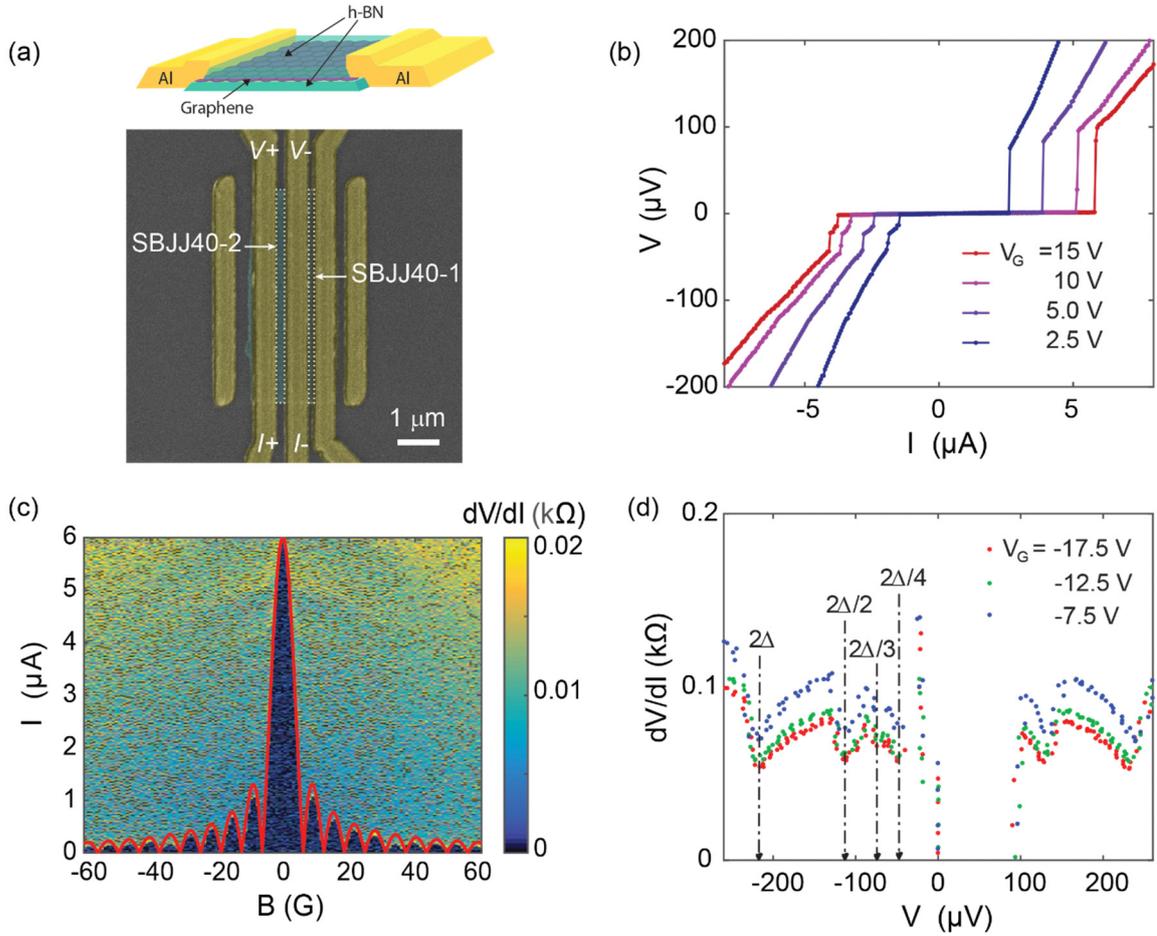

Figure 1. (a) (Upper panel) Schematic diagram of our planar graphene Josephson junctions (pGJJs) with boron nitride (BN)-encapsulated graphene contacted by Al electrodes. (Lower panel) False-colored scanning electron microscopy (SEM) image of two pGJJs: SBJJ40-1 (channel length: ~120 nm) and SBJJ40-2 (channel length: ~220 nm). (b) Current–voltage (I–V) characteristic curves of SBJJ40-2 for different gate voltages $V_G$. Bias current was swept from negative to positive. (c) Differential resistance map as a function of magnetic field and bias current; the Fraunhofer pattern. (d) Bias spectroscopy with dips in the differential resistance of the junction arising from multiple Andreev reflection.

All of the data in this letter, except for the $T$ dependence, were taken at the base temperature of 15 mK. Figure 1(b) shows typical Josephson behavior of the current–voltage

($I$–$V$) curves for SBJJ40-2, with sharp switching to the resistive state at different values of $I_c$ for varying gate voltages ($V_G$). The regular magnetic-field modulation of $I_c$ (Fraunhofer pattern; Fig. 1(c)), satisfying the relation $I_c = I_0 \left| \frac{\sin \pi \Phi / \Phi_0}{\pi \Phi / \Phi_0} \right|$, corresponds to a uniform current distribution along the width of the junction. Here, $I_0$ is the critical current in zero field, $\Phi$ is the perpendicular magnetic flux threading the junction area, and $\Phi_0$ is the flux quantum. The $B$-field period of the oscillation ~6.2 G yields $L + 2\lambda \sim 710$ nm, where the field penetration depth $\lambda$ (~250 nm) is in good accord with previous reports [26,27]. Fig. 1(d) shows the measured differential resistance as a function of the bias current; clear dips appeared via multiple Andreev reflection, leading to an estimation of $\Delta_0$ (~110 μeV) by matching the dip positions with $2\Delta_0 / n$ for the best-fit integer values of $n$.

Fig. 2(a) shows $R_N$ measured at 4.2 K for varying $V_G$. The asymmetry of the resistance between negative and positive sides of $V_G$ is due to electron doping of the graphene layer near the electrode contact induced by the Fermi level mismatch. This leads to a junction resistance at positive $V_G$ that is significantly lower than that of the negative side, where a less transmitting n-p-n junction forms. The Fabry–Perot oscillation observed at negative $V_G$, due to the interference of reflected carriers at the two p-n boundaries, indicates the ballistic transport in the graphene insert. Comparing $R_N$ at positive $V_G$ with the ballistic-limit value of $R_Q = \frac{h}{4e^2} \cdot \frac{1}{N}$ leads to a contact resistivity of ~30 Ω μm, which corresponds to a total transmission probability of $\tau = R_Q / R_N > 0.70$ for $V_G > \sim 2$ V. Here, $N \left( = \frac{2W}{\lambda_F} \right)$ is the number of carrier propagation modes for the channel width $W$ and the Fermi wavelength $\lambda_F$. Fig. 2(b) shows the $dV/dI$ map of SBJJ40-2 as a function of the bias and $V_G$, where the boundary of the dark blue region represents $I_c(V_G)$ of the junction. $I_c(V_G)$ also exhibited asymmetry across the Dirac point, similar to the one in $(R_N)^{-1}$. $I_c$ exceeded ~6 μA at the highly transmitting n-doped region, whereas it was reduced to ~1 μA at the p-doped region. We emphasize that $I_c$ in our device is comparable to those of previous reports on pGJJs of similar widths [13,14]. However, as the junctions consisted of Nb [13] and MoRe [14] superconducting electrodes with gap sizes larger than Al by an order, the normalized coupling strength of our junctions is significantly higher than in those previous reports.

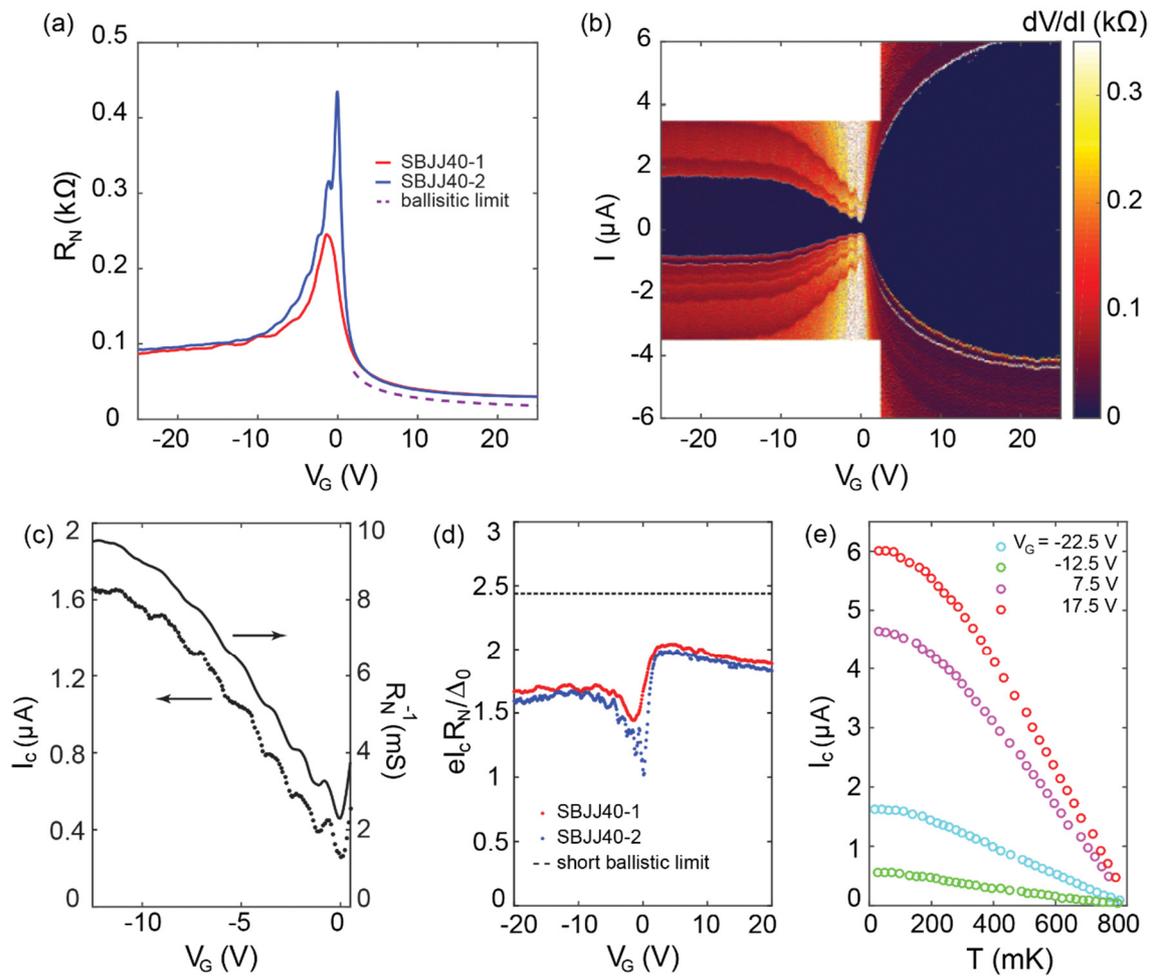

Figure 2. (a) Normal-state junction resistance $R_N$ of SBJJ40-1 and SBJJ40-2 measured at 4.2 K for varying $V_G$. The dashed line corresponds to the ballistic limit with perfect transmission ($\tau = 1$). (b) Differential resistance map of SBJJ40-2 as a function of $I$ and $V_G$. (c) In-phase Fabry–Perot oscillation of the junction critical current $I_c$ and the normal-state conductance $(R_N)^{-1}$ of SBJJ40-2, which indicates ballistic pair transport. (d) The $I_c R_N$ product

normalized by $\Delta_0/e$ of SBJJ40-1 and SBJJ40-2 for varying $V_G$. The dotted line is the theoretical limit, ~2.4, for a short ballistic (SB) pGJJ. (e) $I_c(T)$ of SBJJ40-2 for various $V_G$.

We confirmed the ballistic nature of the Josephson coupling by the oscillation of $I_c$ with $V_G$, which is in phase with the oscillation of $(R_N)^{-1}$ due to the Fabry–Perot interference (Fig. 2(c)). This provides direct evidence for ballistic Josephson coupling, which is mediated by the coherent transport of ballistic quasiparticles via the Andreev bound channel in graphene. Similar ballistic pair transport has been reported in pGJJs [13,14,16,17,28]. The coherence length in our pGJJs was estimated to be $\xi = \hbar v_F/2\Delta_0$ ~2.3 μm, where $\hbar$ is the Planck's constant divided by $2\pi$ and $v_F$ is the Fermi velocity in graphene. Based on the channel lengths of our junctions (~120 nm for SBJJ40-1 and ~220 nm for SBJJ40-2), which are much shorter than the estimated value of $\xi$, our junction should be in the short-junction regime.

Figure 2(d) shows the $I_c R_N$ product (normalized by $\Delta_0/e$) vs $V_G$ for the two pGJJs, the values of which differed by only ~ 5% under sufficiently doped conditions, although $L$ of the two junctions differed by a factor of two. The insensitivity of the $I_c R_N$ product to $L$ strongly indicates that our junctions were in the short-junction regime; had they been in the long-junction limit, the shorter junction (SBJJ40-1) would have had a value of $eI_c R_N/\Delta_0$, two-fold larger than that of the longer one (SBJJ40-2), as the $I_c R_N$ product for a long ballistic junction is inversely proportional to $L$. In addition, $eI_c R_N/\Delta_0$ value itself, reaching up to ~ 2 in our devices, is close to the theoretical prediction (~2.4) for the SB Josephson coupling in pGJJs [20]. It is an order of magnitude larger than the previous report claiming for the SB Josephson coupling in pGJJs [14] and twice as large as that of a diffusive pGJJ with Al electrodes [9]. This value corresponds to the strongest Josephson coupling among the planar proximity JJs studied to date, including JJs based not only on graphene (~1.0) [9] but also on other insert materials such as semiconducting nanowires (~1.1) [29] and the two-dimensional electron gas system (~0.9) [30].

We examined the junction characteristics based on the $T$ dependence of $I_c$. $I_c(T)$ curves of SBJJ40-2 for positive $V_G$ (7.5 and 17.5 V) in Fig. 2(e) are monotonically convex upward ($d^2I_c/dT^2 < 0$). However, the trend becomes marginal for $V_G = -12.5$ V and even slightly convex downward ($d^2I_c/dT^2 > 0$) near $T_c$ for $V_G = -22.5$ V. According to the KO theory, $I_c(T)$ becomes monotonically convex upward only for a short junction. $I_c(T)$ for a long junction should show an exponentially decaying tail near $T_c$ ($\approx 860 \pm 20$ mK for SBJJ40-1 and SBJJ40-2), corresponding to a clearly convex downward behavior, which is qualitatively different from the one for $V_G = -22.5$ V.

Conventionally, the KO theory is used to examine short Josephson coupling characteristics in a proximity JJ. The KO theory, which is valid for a point junction, however, cannot accurately describe the behavior of a wide junction, as used in this study. It is also required to take account of the carrier doping occurring near the graphene-superconductor contact. Thus, we adopt the theoretical model by Takane and Imura (TI model) proposed recently for SB pGJJs [21,22]. This model includes the effect of the induced carrier inhomogeneity in graphene and the transparency of the graphene-superconductor interfaces, where $I_c(T)$ may become convex downward even for a pGJJ in the SB regime.

Due to the electronic carrier doping near the graphene-superconductor contact, an electron experiences a potential well and is refracted at the interface. This feature is included by adopting the electrochemical potential shown in the lower panel of Fig. 3(a), where $\mu$ is the undoped value in graphene and $U$ is its variation due to the carrier doping. As the refraction angle and the transmission probability depend on $U/\mu$, it also affects $I_c(T)$ of a pGJJ in the SB regime. Another feature of this model is that the tunneling of carriers between a graphene sheet and superconducting electrodes is explicitly characterized by using the normalized coupling strength $r$. The Josephson current and its critical value in the SB regime are obtained by applying a quasiclassical thermal Green's function approach to the TI model (refer to Supplemental Material and Ref. [22] for details). It is noted that the TI model deals with the n-n'-n-type unipolar carrier doping in graphene, which can be modified for the n-p-n-type bipolar doping [15].

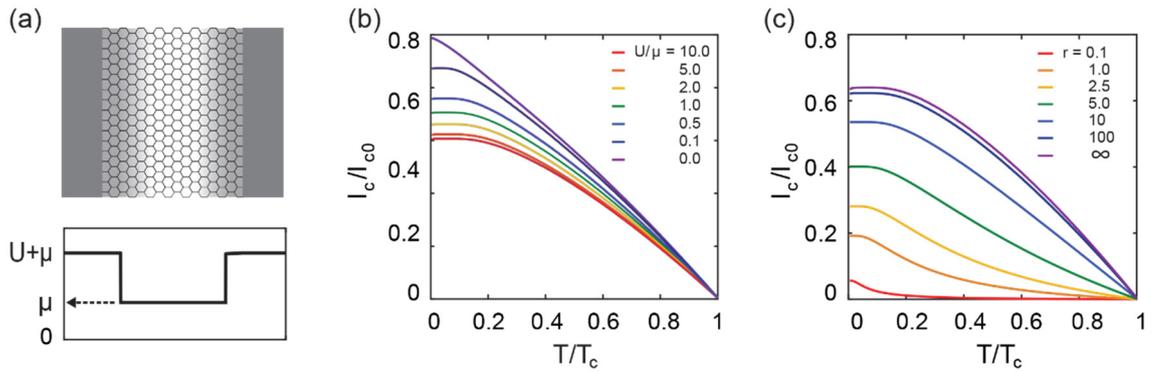

Figure 3 (a) (upper) Schematic diagram of the spatial distribution of the electrochemical potential in a pGJJ and (lower) the simplified electrochemical potential distribution used in the calculation. The arrow indicates the undoped electrochemical potential in graphene. (b) $U/\mu$ dependence of $I_c(T)$ for $r=100$. (c) $I_c(T)$ for different values of $r$ between 0.1 and infinity for $U/\mu=5$, which corresponds to $V_G \sim 10$ V in our experiment.

Fig. 3(b) shows a set of $I_c(T)$ curves, in the SB regime of the TI model, for different values of $U/\mu$ in the range between 0 and 10, for the coupling strength $r$ fixed at 100. The overall shape of the curves were relatively insensitive to the variation of $U/\mu$, although $I_c(T=0)$ continued to decrease with increasing $U/\mu$. However, varying $r$ for a fixed $U/\mu$ at 5 resulted in steep changes in the shape of $I_c(T)$, along with a reduction of $I_c(T=0)$ (Fig. 3(c)). For $r<10$, $I_c(T)$ became convex downward approaching $T_c$ from below, whereas it was monotonically convex upward for $r>10$.

Figs. 4(a) and (b) show the representative fitting of $I_c(T)$ for SBJJ40-2 (at $V_G=7.5$ V) and SBJJ40-1 (at $V_G=10.5$ V) with both the TI and KO models. Both data sets showed convex-upward $T$ dependences, which seem to fit with the regular SB JJ behavior of the KO model. However, the detailed $T$ dependence deviated significantly from the KO prediction. Here, the KO curves were generated for the transmission probability of 0.75 (SBJJ40-1) and 0.72 (SBJJ40-2), the values estimated from the $R_Q/R_N$ ratio of the two junctions (see Fig. 2(a)). On the other hand, the data sets were well fit by the TI model. In the fitting, $U$ (= 0.6 eV) was taken to be the work function difference of graphene and titanium, and $\mu$ was determined for a given $V_G$, with $r$ as the best-fit parameter. The fit led to $r$ ~100 (notably, the fitting is insensitive to $r$ larger than this value), implying a strong coupling between the Al electrode and graphene layer.

We adopted another pGJJ (SBJJ33), which had a channel length ~200 nm similar to SBJJ40-2 (~220 nm), again far shorter than $\xi$. The junction also showed the ballisticity with the Fabry-Perot oscillation both in $(R_N)^{-1}$ and $I_c$ (see Fig. S4), whereas $eI_cR_N/\Delta_0 \sim 1.0$, which is half the value of SBJJ40-2 yet comparable to or larger than the values of any pGJJs reported previously. $I_c(T)$ curves of this junction changed from convex upward near $T=0$ to convex downward with increasing temperature near $T_c$ ($\approx 680$ mK for SBJJ33) (Fig. 4(c)), which is similar to the behavior reported earlier in Ref. [14] (see also Fig. S5). This does not fit with the conventional theory, either for a short junction with monotonically convex upward $I_c(T)$ or for a long junction with an exponentially decaying tail near $T_c$. In striking contrast, however, the curves show a near-perfect fit with the TI model over the entire temperature range below $T_c$ with a low coupling strength of $r \sim 1.1$, which tends to increase slightly with $V_G$ (see the inset of Fig. 4(c)). As the parameter $r$ governs the transmission through the doped graphene–superconductor contacts, this moderate increase of $r$ with $V_G$ is due to the reduction of the Fermi-level mismatch. In fact, it is obvious that SBJJ33 was in a short-junction limit as $\Delta_0$ (~115 $\mu$eV) was much smaller than the Thouless energy $E_{th}$ (~ 2.6 meV).

$I_c(T)$ curves of all our pGJJs fit extremely well with the SB behavior of the TI model in the entire temperature range below $T_c$. While the KO model is valid for a one-dimensional point-contact JJ, our pGJJs have an additional momentum degree of freedom in the direction perpendicular to the Josephson current. The spatial carrier inhomogeneity near the superconducting contacts activated this additional degree of freedom, which reduced $I_c$ with a deviation of its temperature dependence from the KO prediction. Although our junctions consisted of identical contact material in a fully ballistic regime, the junction transmission probability $\tau$ differed by ~15% between SBJJ40's and SBJJ33 devices, whereas the two junctions, SBJJ40-1 and SBJJ40-2, which were fabricated simultaneously, showed only a slight difference in τ. Contact quality was determined mostly during the e-beam deposition process, which not only affects the normal-state transmission but also makes a drastic change in $I_c$ and its $T$ dependence, as predicted by the TI model.

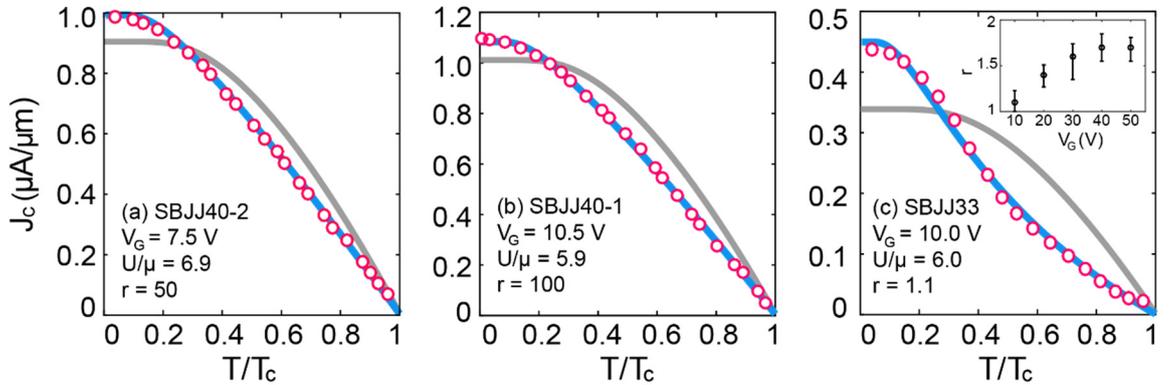

Figure 4. $I_c(T)$ for three pGJJs; SBJJ40-2, SBJJ40-1, and SBJJ33. Void circles represent data. Fits with the Takane–Imura (TI) and Kulik–Omel'yanchuk (KO) models are shown in blue and gray, respectively. Here, the KO curves are generated for the transmission probabilities of 0.75, 0.72, and 0.64 for the three devices, respectively.

In conclusion, the observed large $I_c R_N$ product in our pGJJs in comparison with $\Delta_0/e$ points to SB Josephson coupling. The temperature dependence of $I_c$ of the pGJJs deviated considerably from the conventional KO model. We demonstrate that $I_c(T)$ was well fit by the TI model, which correctly takes into account the effect of carrier transmission via the inhomogeneously doped graphene layer. We found that the detailed $I_c(T)$ is sensitive to the effect of carrier doping, which is represented by the parameter $U/\mu$ in the TI model. The shape of $I_c(T)$ near $T_c$ is, however, mainly determined by the transparency of the graphene-superconductor contacts represented by the parameter $r$; e.g., $I_c(T)$ shows a convex downward dependence in the regime of a relatively small $r$, which is yet different from the conventional long-junction behavior. To summarize, we have identified the two crucial elements for characterizing a pGJJ: the effect of carrier doping near the superconducting contacts and the quality of the graphene-superconductor interfaces. This key finding enables correct characterization of SB strong Josephson coupling, which is essential for generating gate-tunable quantum states with high coherence.

This work was supported by the National Research Foundation of Korea (NRF) through the SRC Center for Topological Matter, POSTECH, Korea (Grant No. 2011-0030046 for H.-J.L.), the SRC Center for Quantum Coherence in Condensed Matter, KAIST, Korea (Grant No. 2016R1A5A1008184 for G.-H.L.), KAKENHI (Grant Nos. 15K05130, 15K05131, and 15H03700 for Y.T. and K.-I.I.), and the Elemental Strategy Initiative conducted by the MEXT and JSPS KAKENHI, Japan (Grant Numbers JP26248061, JP15K21722, and JP25106006 for K.W. and T.T.).

To whom correspondence should be addressed. E-mail: hjlee@postech.ac.kr (H.-J. Lee)

# Supplemental Material

# Short Ballistic Josephson Coupling in Planar Graphene Junctions with Inhomogeneous Carrier Doping

Jinho Park[1], Jae Hyeong Lee[1], Gil-Ho Lee[1], Yositake Takane[2], Ken-Ichiro Imura[2], Takashi Taniguchi[3], Kenji Watanabe[3], Hu-Jong Lee[1]

[1]Department of Physics, Pohang University of Science and Technology, 37673, Korea
[2]Department of Quantum Matter, AdSM, Hiroshima University, 739-8530, Japan
[3]Advanced Materials Laboratory, National Institute for Materials Science, 305-0044, Japan

To whom correspondence should be addressed. E-mail: hjlee@postech.ac.kr (H.-J. Lee)

## S1. Device information for SBJJ40-1

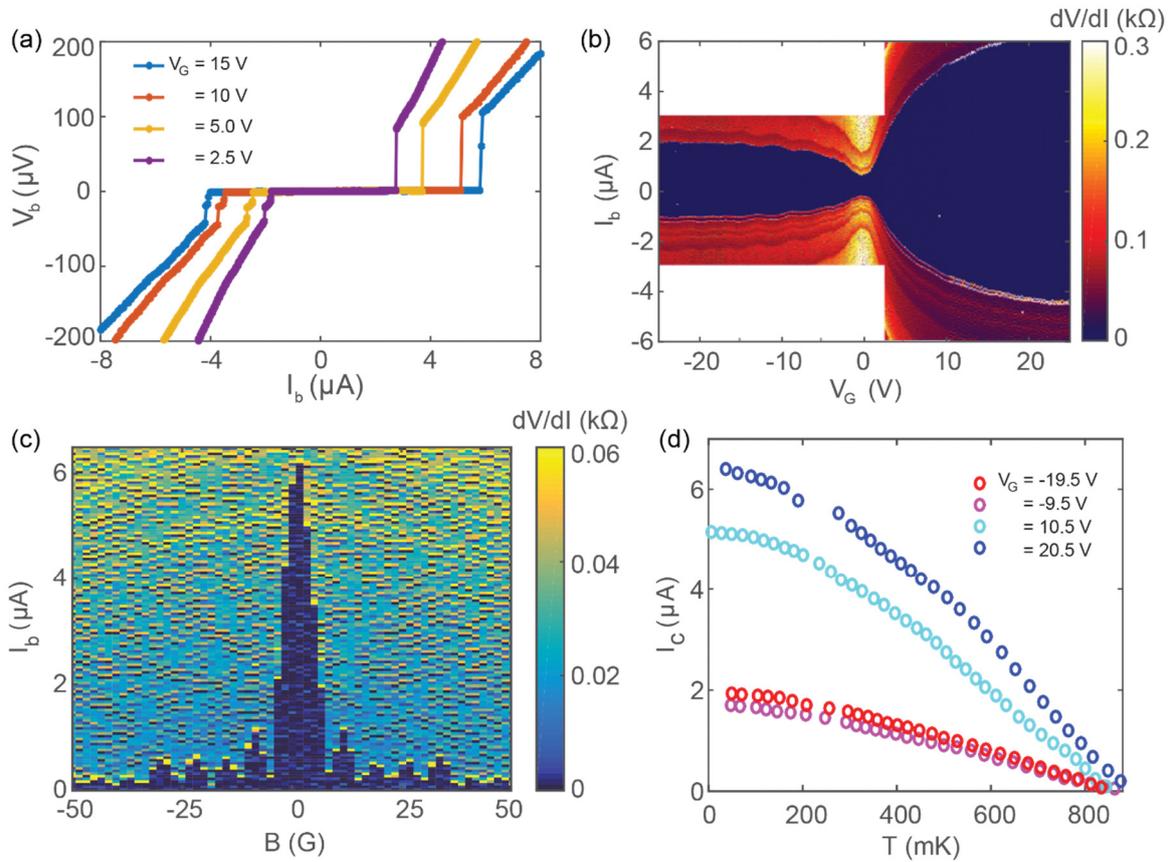

**Figure S5.** (a) *I–V* characteristic curves of SBJJ40-1 for different gate voltages. (b) Color map of the differential resistance of SBJJ40-2 as a function of the current bias and gate voltage. The dark blue region represents the supercurrent range of the junction. (c) Differential resistance map as a function of magnetic field and bias current of the junction (Fraunhofer pattern). (d) Temperature dependence of $I_c$ of SBJJ40-1 for different gate voltages.

The junction SBJJ40-1 exhibits the behavior similar to SBJJ40-2. Sharp switching to the normal state takes place at $I_c$ as shown in Fig. S1(a). $I_c$ of SBJJ40-1 is slightly larger than that of SBJJ40-2, while $R_N$ of two devices in the n-doped region is almost the same. The *B*-field oscillation shown in Fig. S1(c) has the period of ~7.5 G, which coincides well with the junction geometry, with the same penetration depth (~250 nm) as that of SBJJ40-2.

## S2. Simulation results for $I_c(T)$ in TI model in comparison with KO model

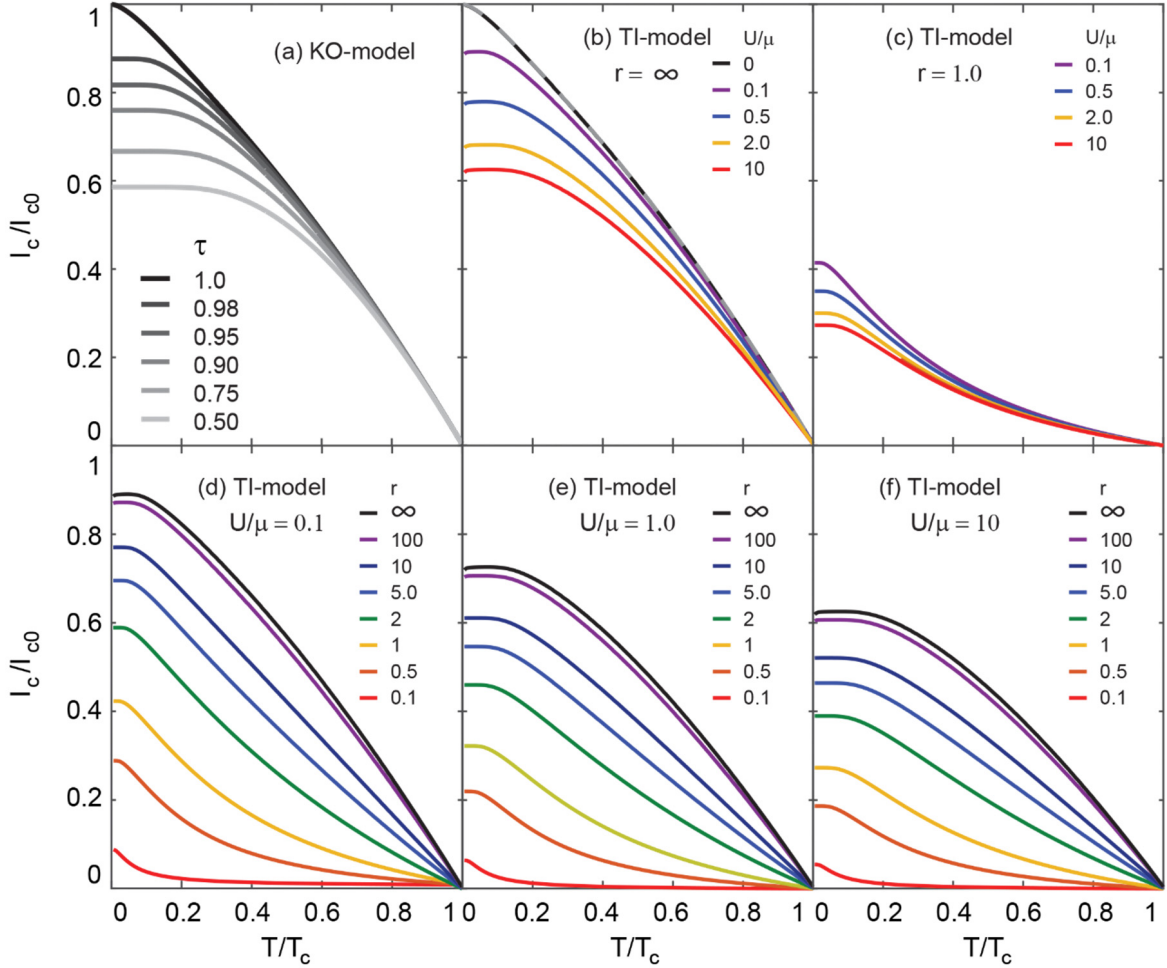

**Figure S6.** (a) $I_c(T)$ curves generated by the KO model, for various transmission probability $\tau$ of a short ballistic Josephson junction. (b), (c) The $U/\mu$ dependence of $I_c(T)$ curves for fixed values of $r=\infty$ and $r=1$, respectively. (d), (e), (f) The $r$ dependence of $I_c(T)$ curves for fixed values of $U/\mu=0$, 1, and 10, respectively.

We discuss the TI model [1] in more detail for various parameter values of $U/\mu$ and $r$ in comparison with the KO prediction [2]. Fig. S2 (a) shows the curves, generated by the following Eq. (S1), of the conventional KO model, which is widely used for examining the short ballistic junction characteristics:

$$I_J = \frac{eN\Delta(T)}{\hbar} \frac{\tau \sin\phi}{\sqrt{1-\tau\sin^2\frac{\phi}{2}}} \tanh\left(\frac{\Delta(T)}{2k_BT}\sqrt{1-\tau\sin^2\frac{\phi}{2}}\right), \quad (S1)$$

where $\Delta(T)$ follows the temperature dependence of BCS theory. $N$ is the number of carrier propagation modes. Here, the transmission probability $\tau$ is the single fit parameter, which can be regarded as the total transmission probability in the normal-junction state. The curve shows monotonically upward regardless of the value of $\tau$. All the curves plotted in Fig. S2 are normalized by $I_{c0}$, the maximal value of critical current at $T=0$ corresponding to $\tau=1$, which gives the value of $I_{c0}R_Q = \pi\Delta_0/e$.

We briefly introduce the TI model for SB pGJJs before we discuss the result of the TI curves simulation. Applying a quasiclassical thermal Green's function approach to the TI model described in the main text, the Josephson current in the SB regime is expressed as

$$I_J(\phi) = I_{c0} \int_0^{\frac{\pi}{2}} d\varphi \cos\varphi \, \frac{k_BT}{\Delta_0}$$
$$\times \sum_{\omega>0} \frac{4\tilde{\Delta}^2(r)\cos^2\theta\cos^2\varphi\sin\phi}{\tilde{\omega}^2(r)\cos^2\theta\cos^2\varphi + \Omega^2(r)(1-\sin\theta\sin\varphi)^2 - \Omega^2(r)(\sin\varphi-\sin\theta)^2\cos 2kL + \tilde{\Delta}^2(r)\cos^2\theta\cos^2\varphi\cos\phi},$$

(S2)

where $\omega$ is the fermion Matsubara frequency, $\varphi$ and $\theta$ are respectively the incident and refraction angles, and $I_{c0}$ corresponds to the critical current at $T=0$ in the KO theory for perfect transmission ($\tau=1$). Here, $\tilde{\omega}$ and $\tilde{\Delta}$ as well as $\Omega = \sqrt{\tilde{\Delta}^2 + \tilde{\omega}^2}$ depend on $r$ as $\tilde{\omega} = [1+\eta(r)]\omega$ and $\tilde{\Delta} = \eta(r)\Delta$ with $\eta(r) = \frac{r\Delta_0}{\sqrt{\Delta^2+\omega^2}}$, and $\theta$ is determined by $U/\mu$ for a given $\varphi$. The critical current is obtained by $I_c = \max\{I_J(\phi)\}_\phi$. The integral gives a value between 0 and 1 with the variation of $U/\mu$, $r$, and $T$ (refer to Ref. [2] for more details).

The $I_c(T)$ curves generated by the TI model are presented in Figs. S2(b) ~ (f), also normalized by $I_{c0}$. As the junction coupling strength represented by the parameter $r$ becomes infinity, the curve becomes monotonically convex upward regardless of $U/\mu$ [Fig.

S2(b)]. The curves for $r=1$, however, become convex downward in most temperature range below $T_c$, which conflicts with the KO model. The TI curve with $U/\mu=0$ and $r=\infty$ is identical to the KO-curve with perfect transmission $\tau=1$. $I_c(T)$ curves at different values of $r$ shows the following general tendency, which is insensitive to the values of $U/\mu=0.1$, 1, and 10. Monotonically convex upward curves appear for $r>100$, while for $r<10$ $I_c(T)$ changes from convex upward to convex downward approaching $T_c$ from below. With decreasing $U/\mu$, $I_c(T=0)$ keeps decreasing.

## S3. Device information for SBJJ33

The normal-state junction resistance $R_N$ measured at 4.2 K for varying $V_G$ is shown in Fig. S3(a). The resistance for positive and negative $V_G$ is highly asymmetric. The total transmission probability of $\tau = R_Q/R_N$ (> 0.60) for $V_G$ sufficiently away from the Dirac point indicates that the junction of SBJJ33 is less transmissive than SBJJ40-2. The junction also shows clear oscillation of $R_N$ for negative gate voltages due to the formation of an n-p-n junction.

For the superconducting state, the junction shows an abrupt switching at $I_c$ to the normal state in its $I-V$ characteristics. The critical current density, smaller than 0.5 μA/μm at 10 V, is about a half of that of SBJJ40's, while the normal-state transmission was not much different from each other. With $\Delta_0 \sim 115$ $\mu$V, which is almost the same as that of SBJJ40's, the value of $eI_cR_N/\Delta_0$ is almost a half of the value of SBJJ40's. As presented in Fig. S3(f), the magnetic field dependence of $I_c$ in SBJJ33 also exhibits a clear Fraunhofer-like pattern with the period of 38 G, corresponding to $L+2\lambda \sim 600$ nm.

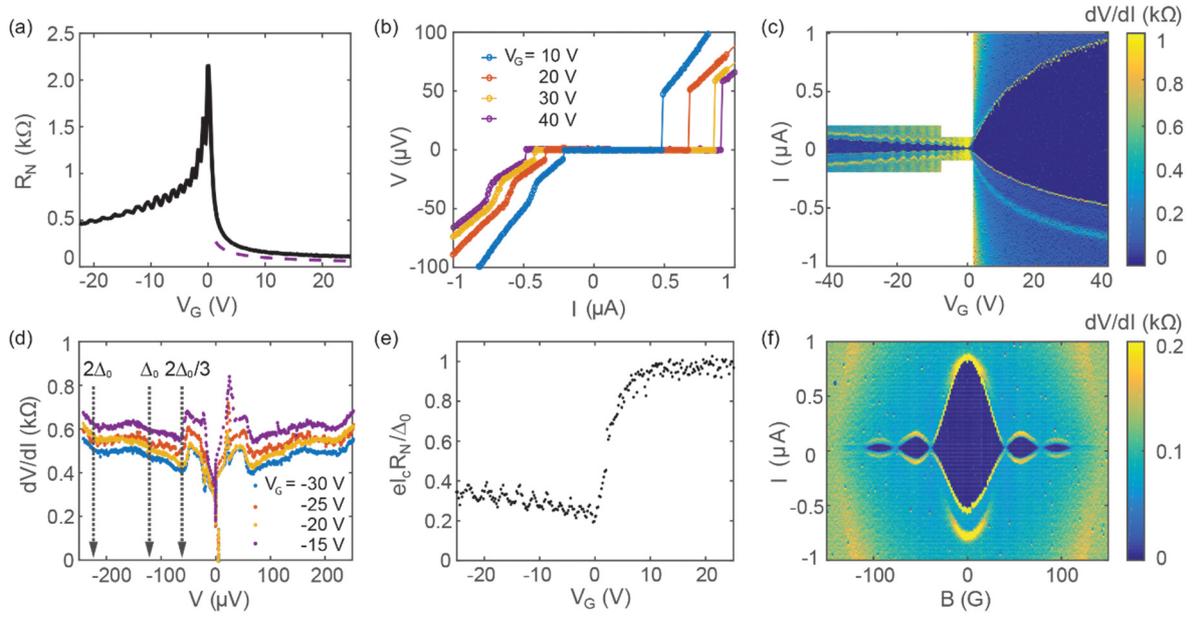

**Figure S7.** (a) The normal-state resistance versus gate voltage for SBJJ33 with the quantum ballistic limit plotted by a dashed purple line. (b) *I-V* characteristic curves of SBJJ33 for positive gate voltages. (c) Color map of the differential resistance of SBJJ33 as a function of the bias current and gate voltage. The dark blue region represents the supercurrent range of the junction. (d) Differential resistance versus bias voltage shows the peaks and dips, originated from the multiple Andreev reflection. The dip positions are used to estimate $\Delta_0$. (e) $eI_cR_N/\Delta_0$ for various gate voltages. The value of $eI_cR_N/\Delta_0$ reaches up to ~1 for positive gate voltages sufficiently away from the Dirac point. (f) Differential resistance map as a function of magnetic field and bias current of the junction (Fraunhofer pattern).

## S4. Ballisticity of SBJJ33

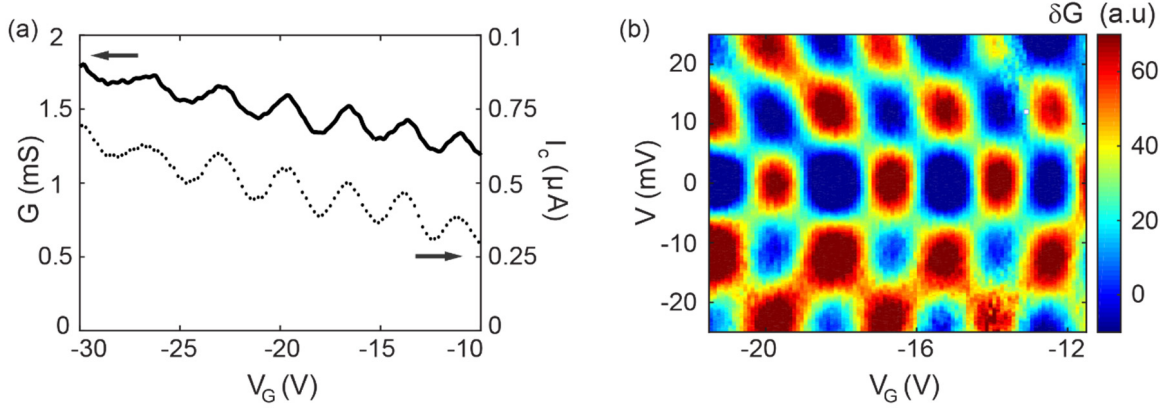

**Figure S8.** (a) Periodic oscillation of $I_c$ with $V_G$, which is in phase with the normal-state conductance $G\,(=R_N^{-1})$. (b) Color map of oscillation of conductance for the function of bias voltage and gate voltage.

The SBJJ33 also exhibits ballisticity with clear Fabry-Perot-type oscillations in both normal-state conductance and $I_c$ in the Josephson-coupled state. Similar result has been reported in studies of ballistic Josephson coupling with planar graphene junctions [3-5]. The length of cavity described in Fig. 3(a) of the main text was estimated from the resonance condition $2L^* = m\lambda_F$, where $L^*$ is the length of the cavity (the region of graphene unaffected by the electrode doping), $m$ is an integer, and $\lambda_F$ is the Fermi wavelength. We obtained $L^* \sim 170$ nm from the gate voltage difference between the adjacent peaks of $I_c$ or $G$ shown in Fig. S4(a).

In addition, from the dispersion relation $\varepsilon_F = hv_F / \lambda_F$, the energy scale of the resonance $\varepsilon_0$ is expressed as $hv_F / 2L^*$. This indicates the oscillatory behavior can also appear as a function of the bias voltage applied across the junction. Thus, the oscillation of the conductance is governed by two independent variables, $V_G$ and $V$, which results in a checkerboard pattern as shown in Fig. S4(b). The channel length estimated by the period of the bias voltage (~24 mV) also gives $L^* \sim 160$ nm, similar to the one determined from the oscillation period of $V_G$ as given above.

## S5. Data sets for other gate voltages

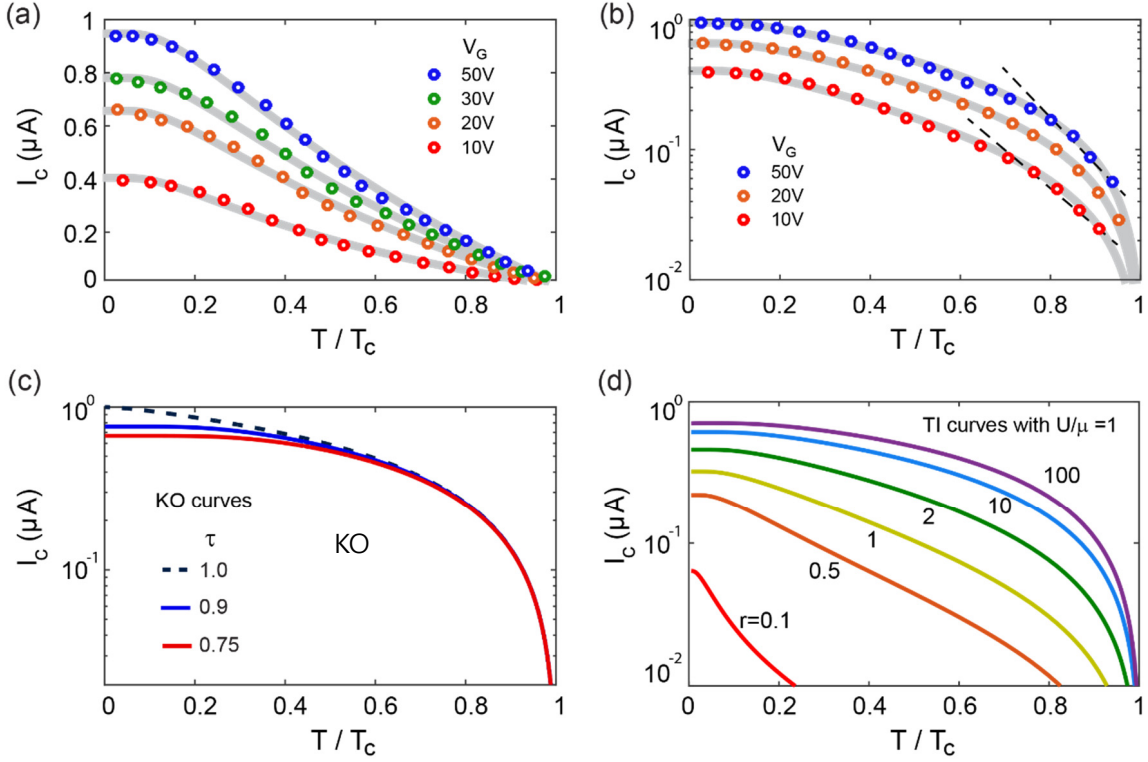

**Figure S9.** (a) Linear-scale fitting of $I_c(T)$ curves for SBJJ33 with the TI model for different gate voltages. All the curves fit very well with relatively low value of the junction coupling strength $r\ (\approx 1-1.8)$. (b) Corresponding semi-logarithmic plot of the data sets for $V_G = 10, 20, 50$ V to examine the exponential decaying of $I_c(T)$ approaching $T_c$ from below. (c, d) The KO curves and TI curves plotted in a semi-logarithmic scale. (c) KO curves are plotted for the transmission probability of 0.9 and 0.75. (d) TI curves with $U/\mu = 1.0$ are plotted for various $r$ between 0.1 and 100.

We plot the $I_c(T)$ curves of SBJJ33 for various $V_G$ from 10 to 50 V. Although $I_c(T)$ turns convex downward below $T_c$, all the data for different values of $V_G$ fit well with the short ballistic behavior of the TI model, which does not fit with the conventional KO theory. We also examine the possibility that our junction is in long junction regime. $I_c(T)$ curves of SBJJ33 in Fig. S5(a) are replotted in a semi-logarithmic scale to examine its exponential decaying behavior below $T_c$ as $I_c(T) \sim I_{c0} \exp(-T/T_{0b})$, the typical characteristics of a long junction with $I_c$ decreasing linearly with $T$ in a semi-logarithmic plot. Here, $T_{0b}$ is the temperature corresponding to $k_B T_{0b} = E_{th} = hv_F / L$. However, in contrast to the exponentially decaying of $I_c$ with $T$ the curves shown in Fig. S5(b) have a monotonically convex upward behavior. The linear broken guide lines near $T_c$ corresponds to $T_{0b} \sim 100$ mK, which is far below the value of $T_{0b}$ ($\sim 30$ K, estimated for SBJJ33 with $L \sim 200$ nm).

In Fig. S5(c), we present the KO curve plotted in a semi-logarithmic scale, for different junction transmission probability, to compare the fits with the short-junction behavior claimed previously in Ref. 5. It shows high discrepancy with the fitting provided in Fig. 2 of Ref. 5. Although the curves in Fig. S5(b) appear similar to the KO-model curves shown in Fig. S5(c) in a semi-logarithmic plot, they represent convex-downward $I_c(T)$ as shown in Fig. S5(a). In fact, we notice that even convex-upward curves, in Fig. 2 of Ref. 5, in a semi-logarithmic plot are convex downward in a linear plot, which cannot be fit with the short-junction KO curves. In this sense, short-junction characteristics should be judged in a linear plot of $I_c(T)$. Here, we point out that the data shown in Ref. 5 can be fit with the TI model with the relatively small $r$ close to 1 (the yellow curve in Fig. S5(d)) if a sufficiently small scaling factor is adopted. This small value of $r$, however, is in contradiction with its high transmission of up to ~0.9. In addition, the small value of $I_c R_N$ product $\sim 0.3\Delta_0 / e$ also remains an unsolved feature.